\documentclass[preprint,amsmath,amssymb,nofootinbib]{revtex4}
\pdfoutput=1

\usepackage{graphicx}
\usepackage{url}
\usepackage{color}
\usepackage{xspace}
\usepackage{float}
\usepackage{hyperref}
\usepackage{ulem}
\usepackage{setspace}
\usepackage{amssymb}
\usepackage{amsmath}
\usepackage{natbib}
\usepackage{enumitem}
\usepackage{multirow}

\newcommand{\nlox}{\textnormal{\textsc{NLOX}}\xspace}

\newcommand{\tred}{\textnormal{\texttt{TRed}}\xspace}

\newcommand{\fortran}{\textnormal{\texttt{Fortran}}\xspace}

\newcommand{\nlo}{next-to-lowest-order\xspace}

\begin{document}          

\title{Updates to the One-loop Provider \nlox}

\author{D.~Figueroa}
\email{daf14f@my.fsu.edu}
\affiliation{Physics Department, Florida State University,
Tallahassee, FL 32306-4350, U.S.A.}
\author{S.~Quackenbush}
\email{squackenbush@hep.fsu.edu}
\affiliation{Physics Department, Florida State University,
Tallahassee, FL 32306-4350, U.S.A.}
\author{L.~Reina}
\email{reina@hep.fsu.edu}
\affiliation{Physics Department, Florida State University,
Tallahassee, FL 32306-4350, U.S.A.}
\author{C.~Reuschle}
\email{christian.reuschle@thep.lu.se}
\affiliation{Department of Astronomy and Theoretical Physics, 
Lund University, SE-223 62 Lund, Sweden}

\begin{abstract}
  In this release note we describe the 1.2 update to \nlox, a computer program for
  calculations in high-energy particle physics. New features since the
  1.0 release and other changes are described, along with usage
  documentation.
\end{abstract}

\maketitle

\section{Introduction}
\label{sec:intro}
\nlox is a one-loop provider that allows for the automated calculation
of one-loop QCD and electroweak (EW) corrections to Standard Model
(SM) scattering amplitudes. Based on Feynman-diagram techniques, it
has been developed to optimize the manipulation of symbolic and tensor
structures recurring in one-loop amplitudes.  The first release of the
code (v1.0)~\cite{Honeywell:2018fcl} consisted of the main
tensor-integral reduction library, \texttt{TRed}, and several
pre-generated processes. Further processes have been produced upon
demand for various projects. All core functionalities of \nlox v1.0 have
been presented in detail in Ref.~\cite{Honeywell:2018fcl} to which we refer.

Since v1.0 the code has seen a substantial number of important additions aimed
at improving the numerical stability of the one-loop amplitudes
and the interface to external codes. Among the most important
additions are improved stability checks in the tensor reduction
performed by \texttt{TRed} and new routines that check at runtime the
IR pole structure of virtual amplitudes against the IR pole structure
of the corresponding real emission, order by order in the QCD+EW
couplings. These new IR checks are implemented using color-correlated
amplitudes that are now also independently obtainable from the
code. Given the impact that these improvements will have on the
succesful use of the code, we have included them in the new NLOX
v1.2 that is being released. 

In the following, Sections~\ref{sec:features} and
\ref{sec:improvements} present a more technical description of all the
code's new features and improvements, while future developments are
outlined in Section~\ref{sec:outlook}.

\section{New features}
\label{sec:features}

\subsection{New modes for perturbative expansions}
\label{sec:expansions}

\nlox now has interface support for automatic compilation of all
sub-amplitudes that contribute to the same overall power of $\alpha_s$
(QCD) and $\alpha_e$ (QED) in the interference between tree level and
tree level or one-loop amplitudes for any given sub-process.  This
\textit{$\alpha$-mode} complement the existing possibility of
compiling sub-amplitudes by specifying the $g$ ($g^2=4\pi\alpha_s$)
and $e$ ($e^2=4\pi\alpha_e$) power of the interfering tree level and
tree level or one-loop amplitudes~\cite{Honeywell:2018fcl}.

The compiled result with a fixed power of both $\alpha_s$ and
$\alpha_e$ can be accessed via the OLP function
\texttt{NLOX\_OLP\_EvalSubProcess} by passing the corresponding string
(\texttt{cp}) specifying the coupling-power combination. In a
\texttt{C++} program this OLP function takes the form:
\begin{center}
\texttt{NLOX\_OLP\_EvalSubProcess(isub, type, cp, p, next, mu, rval, acc)}
\end{center}
where the arguments follow the BLHA standard~\cite{Alioli:2013nda}:
\begin{itemize}
 \item \texttt{isub}: Integer number specifying the index of the desired sub-process as defined in the sub-process list in \texttt{nlox\_process.h}.
 \item \texttt{type}: Character string specifying the interference type, both \texttt{tree\_tree} and \texttt{tree\_loop} are supported.
 \item \texttt{cp}: Character string specifying the coupling-power
   combination of the interference type in question, both
   \texttt{g$i^\prime$e$j^\prime$\_g$i$e$j$} and \texttt{as$X$ae$Y$}
   (where $X=(i^\prime+i)/2$ and $Y=(j^\prime+j)/2$)  are supported.
 \item \texttt{p}: Array of double-precision numbers specifying the external momenta and masses, each momenta must be followed by its mass.
 \item \texttt{next}: Integer number specifying the number of external particles.
 \item \texttt{mu}: Double-precision number specifying the
   renormalization scale (in GeV).
 \item \texttt{rval}: Return value pointer, the container must allow
   for $3$ double-precision numbers to be stored corresponding to
   double pole, single pole, and finite part.
 \item \texttt{acc}: Accuracy pointer, the container must allow for a single double-precision value to be stored. 
\end{itemize}
In a \fortran program this OLP function
takes extra arguments specifying the length of the interference type and of the coupling power combination:
\begin{center}
\texttt{NLOX\_OLP\_EvalSubProcess(isub, type, ltyp, cp, lcp, p, next, mu, rval, acc)}
\end{center}
where the extra arguments of this function are:
\begin{itemize}
 \item \texttt{ltyp}: Integer specifying the length of the interference type string.
 \item \texttt{lcp}: Integer specifying the length of the coupling-power combination string.
\end{itemize}
A detail worth mentioning regarding the usage of the two different 
specifications for coupling powers in NLOX is the following. In the 
$\alpha$-mode, where the character string \texttt{cp} is in the form 
\texttt{as}$X$\texttt{ae}$Y$, all sub-amplitudes with the same overall 
power of $\alpha_s$ and $\alpha_e$ are summed over automatically. In 
contrast to that, when the character string \texttt{cp} is specified in 
the form \texttt{g}$i'$\texttt{e}$j'$\_\texttt{g}$i$\texttt{e}$j$, the 
various sub-amplitudes contributing to the same overall power of 
$\alpha_s$ and $\alpha_e$ can be accessed individually, but must then be 
summed up manually.\\

\subsection{Color-correlated output}
\label{sec:col-cor}

\nlox is now capable of providing
color-correlated Born-level matrix elements $\textbf{B}_{ij}$. The
normalization convention used is:
\begin{equation*}
 \textbf{B}_{ij} = \langle\mathcal{M}^{C_1,...,C_i,...,C_j,...,C_m}|(\textbf{T}^a)_{C_iD_i}(\textbf{T}^a)_{C_jD_j}|\mathcal{M}^{C_1,...,D_i,...,D_j,...,C_m}\rangle
\end{equation*}
where the matrices $\textbf T^a$ denote the generators of the $SU(3)$
algebra and the indices $\{C_i,D_i\}$ are generic indices associated
to the $SU(3)$ representation under which the $i$-th particle
transforms. More specifically~\cite{Catani:1996vz} the
$(\textbf{T}^a)_{C_iD_i}\equiv \textbf{T}_i$ are defined as:
\[ 
  (\textbf{T}^a)_{{C_iD_i}} =
    \begin{cases} 
      (t^a)_{\alpha\beta} & \text{for initial quarks and final anti-quarks}, \\
      -(t^a)_{\beta\alpha} & \text{for initial anti-quarks and final quarks}, \\
      (T^a)_{cb}=if^{cab} &  \text{for gluons}, 
    \end{cases}
\]
where $(t^a)_{\alpha\beta}$ $(\alpha,\beta=1,2,3)$ and
$(T^a)_{cb}=if^{abc}$ ($a,b,c=1,\ldots,8$) are the color-charge
matrices in the fundamental and adjoint representation respectively,
$f^{abc}$ are the $SU(3)$ structure constants, and
$(\textbf{T}_i)^2=C_i$ with $C_i=C_A=N$ if $i$ is a gluon and
$C_i=(N^2-1)/2N$ if $i$ is a quark or an antiquark ($N=3$ for SU(3)).

For direct evaluation, there is a dedicated OLP level function that
calls for the evaluation of color-correlated matrix elements:
  \begin{center}\texttt{NLOX\_OLP\_EvalSubProcess\_CC(isub, type, cp, p, next, mu, borncc, acc)}
  \end{center}
\noindent which can only be called for Born-level matrix elements. 
The container that \texttt{borncc} points to must be able to store the
uncorrelated Born-level matrix element as well as the
$\textbf{B}_{ij}$ components for which $j>i$, while the
components with $i>j$ can be obtained via symmetry from the $\textbf{B}_{ij}$. 
The first element in the container
is reserved for the uncorrelated Born, and the subsequent
$[n(n-1)]/2$ elements correspond to the components of
  $\textbf{B}_{ij}$ using lexicographical order on $(i,j)$.\\

For \fortran programs the OLP function takes extra arguments specifying the length of the interference type as well as the coupling power combination:
\begin{center}
\texttt{NLOX\_OLP\_EvalSubProcess\_CC(isub, type, ltyp, cp, lcp, p, next, mu, rval, acc)}
\end{center}
where the extra arguments of this function are:
\begin{itemize}
 \item \texttt{ltyp}: Integer specifying the length of the interference type string.
 \item \texttt{lcp}: Integer specifying the length of the coupling-power combination string.
\end{itemize}

\subsection{Runtime amplitude-level pole checks}
\label{sec:ir-pole-check}
Routines that calculate the IR pole structure of the real
  emission associated to a given Born amplitude have been newly implemented within \nlox,
but are only computed in the \textit{$\alpha$-mode} of coupling-power
specification. The virtual IR poles are computed independently in the
combination of loop and counterterm diagrams, and must cancel the real
IR poles in the $G_{\mu}$ input scheme.\footnote{ Because \nlox only
  allows nonzero masses for the $b$ and $t$ quarks, EW renormalization
  in the $\alpha(0)$ scheme formally produces IR poles
  arising from massless fermions that remain after combination with
  those arising from real radiation, and must be compensated
  externally. Thus, in the $\alpha(0)$ scheme, the IR pole-checking feature is only useful for
  strictly QCD corrections, while it works for both QCD and EW
  corrections in the $G_\mu$ scheme (\texttt{GmuScheme = true}).}  If
real and virtual IR poles must cancel against each other, checking for
the goodness of zero of their sum can be used to provide a check on
the stability of the tensor reduction algorithm. The user has access
to this new feature via the \texttt{acc} flag of the 
\texttt{NLOX\_OLP\_EvalSubProcess} function. This flag has different behavior
depending on the particular type of \texttt{cp} the user specifies:
\begin{itemize}
 \item \texttt{cp=g$i^\prime$e$j^\prime$\_g$i$e$j$}: \texttt{acc} is either $0$ for success
   or $-1$ for failure, where failure is 
 internally determined by \texttt{TRed} as described in \ref{sec:tred-stab}.
 \item \texttt{cp=as$X$ae$Y$}: \texttt{acc} is $-1$ for a \texttt{TRed}
   failure. In case of a \texttt{TRed} success the \texttt{acc} flag
   returns the accuracy of the virtual IR single pole relative to the
   real IR single pole. 
\end{itemize}
The calculation of the real IR poles is based on the
dipole-subtraction formalism and it requires the internal evaluation
of color-correlated Born matrix elements~\cite{Catani:1996vz}. 

\subsection{Runtime parameter setting}
In order to change physical parameters
at runtime, \nlox now supports the Les Houches~\cite{Alioli:2013nda}
standard function \texttt{OLP\_SetParameter()}: 
\begin{center}\texttt{NLOX\_OLP\_SetParameter(para, re, im, ierr)}
  \end{center}
whose arguments are:
\begin{itemize}
 \item \texttt{para}: Character array, the name of the parameter to be changed. The currently supported names are:
   \begin{itemize}
    \item \texttt{"alpha\_e"}: $\alpha_e$, the QED fine structure constant.
    \item \texttt{"alpha\_s"}: $\alpha_s$, the QCD coupling constant.
    \item \texttt{"mb", "mt", "mH", "mW", "mZ"}: masses of supported massive particles.
    \item \texttt{"wH", "wW", "wZ"}: widths of supported
      particles. 
     At present unstable fermions are not officially supported, and giving a nonzero width to an
     external particle will result in undetermined
     behavior.
    \item \texttt{"nlf"} the number of light quark flavors to be used
   for renormalization purposes (see Appendix A.3 of Ref.~\cite{Honeywell:2018fcl} for
   more details).
    \item \texttt{"nhf"} the number of heavy quark flavors to be used
     for renormalization purposes (see Appendix A.3 of Ref.~\cite{Honeywell:2018fcl} for
   more details).
    \item \texttt{"mu"} the renormalization scale.
   \end{itemize}
   It is recommended that one change these values only as often as necessary. For masses
   and widths in particular, changing the value can result in expensive reallocations and
   computations.
 \item \texttt{re}: Pointer to a double precision number holding the new real part of the value to be set.
 \item \texttt{im}: Pointer to a double precision number. This should be a dummy variable, as \nlox does not currently support setting parameters to complex values through this function.
 \item \texttt{ierr}: Pointer to an integer to store the return flag of the function.
 Set to 1 if successful and 2 if the name is unknown.
\end{itemize}
For \fortran programs the \texttt{SetParameter}
function takes an extra argument:

\begin{center}
\texttt{NLOX\_OLP\_SetParameter(para, lpara, re, im, ierr)}
\end{center}
where the extra argument of this function is:
\begin{itemize}
 \item \texttt{lpara}: Integer specifying the length of the parameter name string.
 \end{itemize}
Derived complex parameters arising in the complex mass scheme~\cite{Denner:2005fg}, if selected, are calculated using the complex masses, which are built from real masses and widths and recomputed when one is set.

\section{Improvements and other changes}
\label{sec:improvements}

\subsection{\tred stability checks}
\label{sec:tred-stab}

Since its original release, the \tred library
has internally checked the stability of the tensor-integral
reduction. Its primary check is to compare the single pole term (the
coefficient $c_1$ in the expansion
$c_2 \epsilon^{-2} + c_1 \epsilon^{-1} + c_0$ of a given
  tensor-integral coefficient) to an internal fast-evaluating
library for coefficients that are infrared finite. This is a direct
cross-check, that allows early correction of individual coefficients
through recalculation at higher precision, and acts as a first-pass
  filter of potential instabilities, and has now been complemented by
  the IR-pole check described in Section~\ref{sec:ir-pole-check}. In
  this new release of NLOX we have improved its effectiveness
  as described in the following.
 
The driver of instability for the primary reduction method,
Passarino-Veltman \cite{Passarino:1978jh}, is unstable matrix
inversion involving the Gram matrix $G_{ij} \equiv 2 p_i \cdot p_j$
for loop internal momenta $p_i$. In particular, the determinant of
this matrix becomes small as it becomes degenerate, and as it appears
explicitly in the denominator of reduction formulae, an imprecise
cancellation in the numerator of the reduction can result in large
incorrect results. To guard against this problem, in previous versions
\tred has compared the size of the Gram determinant to an
appropriate-scaling quantity, $~E^{2(N-1)}$ for $E$ a scale in the
problem. However, one can improve on the sensitivity and specificity
of this test with more information, reducing the failures for
acceptable phase space points and catching others that may have been
problematic. To this end, we have improved the test to more directly
check the instability of the linear system.

A geometric interpretation of a determinant is that it is the volume
of the parallelepiped spanned by the matrix's row (or column)
vectors. As the vectors become linearly dependent, the volume of this
space shrinks to zero, and the space spanned by them loses a
dimension.  Therefore, a direct measure of the smallness of the
determinant compares it to the maximal volume such a space would have
with the same-sized vectors, i.e., if they were orthogonal.  Our new
test compares $\det G_{ij}$ to $\prod_i |\vec{G_i}|$ (where
$\vec{G_i}$ denotes the array of the $G_{ij}$ for fixed $i$). The specificity
of this test has allowed us to pass many configurations that would
previously have been thrown out, even while preventing unhandled
instabilities that could occur in collinear configurations for some
processes.

Since \tred relies on external scalar integral libraries for the
endpoint of its reduction, it is only as good as the precision of
those libraries. In particular, in the small determinant case above,
combinations of scalar integrals appearing may cancel in the limit of
degenerate configurations. While \tred will attempt to do the
reduction in higher precision, if the scalars only match to a lower
precision, this can result in poor behavior regardless. Therefore we
have reconfigured the installation of the default scalar library,
OneLOop~\cite{vanHameren:2010cp}, to install quadruple-precision versions of its functions, and
adjusted \tred to make use of them as appropriate.

\subsection{Code size optimizations}
Expressions resulting from the algebraic construction of the
expression for a process squared amplitude naturally can be large for
complicated processes, especially if many masses are involved or the
theory has non-trivial expressions for interactions, as in the case of the
broken electroweak theory. We have focused on simplifying
expressions, especially finding overall factors, and finding common
structures such that the resulting code size is smaller. These
optimizations can result in a reduction of code size of as much as a
factor of two in extreme situations compared to the version 1.0
release. These optimizations improve runtime speed as well if the
speed is process-cache or memory-speed limited.

Since much of the resulting algebraic data has been removed from code
generation entirely, and placed into process data files, this is the
dominant contribution to a process package size. We have eliminated
some empty files and redundant structures to further reduce the size
of a process package.

\section{Outlook}
\label{sec:outlook}
We have described the important changes to \nlox since its original
release. The main focus has been on numerical stability. We have
reorganized internal integral stability checks for accuracy and
efficiency. We have also added a new stability check by implementing
the IR singular part of the real radiation using the dipole-subtraction
formalism, and exploiting the finiteness of the combined real and
virtual contribution after the respective IR poles cancels at the level of the final answer.

The implementation of IR-pole checks based on the dipole formalism has also exposed a new
functionality to the user, that of Born-level color-correlated
amplitudes, which may be of use to users wishing to interface \nlox to
Monte Carlo simulators requesting these amplitudes.

This release has also seen the implementation of runtime parameter
setting, further implementing the Les Houches Accord \nlo function
interface and making the process of interfacing \nlox to external
programs more convenient.

While this 1.2 release and its predecessor, 1.1, have focused on
stability and ease of use, we anticipate the next releases will focus
on speed and size efficiency, while also continuing to improve the
ease of use and organization of the non-public process archive
generation features so that the full package may be released.

To access the host \texttt{URL} for the \nlox package,
please go to \url{http://www.hep.fsu.edu/~nlox}.

\section{Acknowledgments}
This work has
been supported by the U.S. Department of Energy under grant DE-SC0010102.
C.R. acknowledges support by the European Union’s Horizon 2020
research and innovative programme, under grant agreement No. 668679.
L.R. is grateful to the Aspen Center for Physics for the
hospitality offered while parts of this work were being completed. 
The work performed at the Aspen Center for Physics is supported
in part by the National Science Foundation under grant NSF PHY-1607611.

\bibliographystyle{apsrev}
\bibliography{nlox-beta}

\end{document}